\documentclass[twocolumn]{revtex4}

\usepackage{amsmath}
\usepackage{amssymb}
\usepackage[english]{babel}

\newtheorem{definition}{Definition}
\newtheorem{lemma}{Lemma}
\newtheorem{theorem}{Theorem}
\newcommand{\bra}[1]{\left. \langle #1 \right|}
\newcommand{\ket}[1]{\left| #1 \rangle \right.}
\newcommand{\opav}[1]{\left\langle #1 \right\rangle}
\newcommand{\tth}{\tilde{\theta}}

\begin{document}
\pagenumbering{roman}
\title{Decomposing generalized measurements into continuous stochastic processes}
\author{Martin Varbanov}
\email{varbanov@usc.edu}
\affiliation{Department of Physics and Astronomy, University of Southern California, \\
Los Angeles, CA  90089}

\author{Todd A. Brun}
\email{tbrun@usc.edu}
\affiliation{Communication Sciences Institute, University of Southern California, \\
Los Angeles, CA  90089}

\date{\today}

\begin{abstract}
One of the broadest concepts of measurement in quantum theory is the generalized measurement. Another paradigm of measurement---arising naturally in quantum optics, among other fields---is that of continuous-time measurements, which can be seen as the limit of a consecutive sequence of weak measurements. They are naturally described in terms of stochastic processes, or time-dependent random variables.  We show that any generalized measurement can be decomposed as a sequence of weak measurements with a mathematical limit as a continuous stochastic process. We give an explicit construction for any generalized measurement, and prove that the resulting continuous evolution, in the long-time limit, collapses the state of the quantum system to one of the final states generated by the generalized measurement, being decomposed, with the correct probabilities. A prominent feature of the construction is the presence of a feedback mechanism---the instantaneous choice weak measurement at a given time depends on the outcomes of earlier measurements.  For a generalized measurement with $n$ outcomes, this information is captured by a real $n$-vector on an $n$-simplex, which obeys a simple classical stochastic evolution.
\end{abstract}

\maketitle


\pagenumbering{arabic}
\small

\section{Introduction}

The first and simplest definition of measurement that one learns in a first course on quantum mechanics involves projection operators which sum up to the identity \cite{Peres}. Usually, these projective operators come from a resolution of the identity generated by a particular observable, represented by a Hermitian operator. A generalization is needed when composite systems are considered---a projective measurement on the whole system is not usually a projective measurement on a part of the system. Every {\it generalized} measurement can be performed by adding an ancilla to our given system, doing an appropriate unitary transforming both system and ancilla, followed by a projective measurement on the ancilla \cite{NiCh}. The dimension of the ancilla must be at least equal to the number of outcomes from the measurement and the number of orthogonal projectors on the ancilla.

A projective measurement can be carried out as a sequence of {\it weak} generalized measurements \cite{NiCh}. Here, {\it weak} means that after each step in the sequence, the system is disturbed by a small amount, but yields only a small amount of information. Each step also takes a certain finite amount of time, and thus so will the final strong projective measurement. This picture of a measurement, as a continuous sequence of infinitesimal steps taking a finite amount of time, contrasts with the usual assumptions we make about abstract measurements---that they are instantaneous and thus strong.

While this construction works for projective measurements, it cannot be immediately adapted to strong {\it generalized} measurements.  We wish to construct a measurement procedure that is continuous in time, but that at long times produces the same result as a given strong measurement:  that is, in the end the system is in one of the correct final states with the correct probabilities. We will prove that such a continuous decomposition exists for any generalized measurement, and give an explicit construction.  Such a continuous procedure can serve a variety of purposes.  For certain problems, a continuous description is useful:  we are able to take derivatives and use other analytical tools of calculus \cite{OrBr1, OrBr2}, apply methods from quantum filtering theory and quantum feedback control. On the level of physical reality, it is difficult to argue whether one particular description of a measurement is more fundamental (if any of them is). For much of the time since the discovery of quantum mechanics it made sense to treat measurements as instantaneous because those were the kind of measurements we were able to perform. With more recent advances in quantum optics and atomic physics, where quantum systems can be continuously monitored, new ways of describing the evolution of the system had to be developed---models such as quantum trajectories or decoherent histories \cite{Ca,Wi,GuBo,Hu,Gi,GiPe,Br1,Br2}.  That there are numerous ways to think about measurements in quantum mechanics, which are not necessarily contradictory, but on the contrary are in some sense complementary.

The purpose of this paper is to construct a quantum continuous stochastic process---a family of consecutive weak measurements---which governs the evolution of the state of the quantum system, and for which the final result (at long times) is the same as a that of a specified (strong) generalized measurement.  We will give an explicit construction of such a process, and prove that it does indeed have the correct long-time behavior.  As we will see, in general this requires that the choice of measurements at later times will depend on the measurement results at earlier time, so this continuous measurement procedure must include feedback.

In Section II we first consider the discrete case of successive weak but finite measurements as a good and fairly simple introduction to what happens in the continuous case.  After that, in Section III we derive a stochastic differential equation for the continuous stochastic process implementing a projective measurement.  This construction is simplified by the fact that the measurement operators all commute, and the final states are all orthogonal, but it will show the approach to be followed in the more general case.  We prove that our stochastic process really does generate the results of the generalized measurement as time goes to infinity. Finally, in Section IV we show how the same thing can be done for {\it any} generalized measurement.  This does not require that the measurement operators commute, nor that the final states be orthogonal, and in general requires feedback from the results of earlier measurements to the choice of later measurements.  We give an example, in the case when the measurement operators {\it do} commute, of how generated quantum state diffusion equations can be derived from our equations.  Section V briefly summarizes our results and conclusions.

\section{Repeated weak measurements for projective measurements}

Let's consider a projective measurement on the system. The measurement operators are denoted by $\hat{P}_i$ and they have the following properties
\begin{align}
& \hat{P}_i^{\dagger} = \hat{P}_i = ( \hat{P}_i )^2, \notag \\
& \hat{P}_i \hat{P}_j = \delta_{ij} \hat{P}_i, \label{E:Label07} \\
& \sum_{i=1}^n \hat{P}_i = \hat{I}. \notag
\end{align}
We would also assume that initially our system is in state $\ket{\psi_0}$. After we perform the projective measurement \eqref{E:Label07} on the system, it states collapses to one of the following states
\begin{align}
\ket{\bar{\psi}_i} = \frac{\hat{P}_i \ket{\psi_0}}{\sqrt{p_i^0}} \label{E:Label33}
\end{align}
with probability $p_i^0 = \bra{\psi_0} \hat{P}_i \ket{\psi_0}$ respectively.

We now give the discrete procedure which decomposes the projective measurement into a sequence of weak measurements. We denote by $\Delta^n$ the classical state space
\begin{equation*}
\Delta^n=\left\{ x \in (0,1)^n \subset \mathbb{R}^n \left| \sum_{i=1}^n x^i=1 \right\} \right.
\end{equation*}
and by $\bar{\Delta}^n$ -- the closure of $\Delta^n$
\begin{equation*}
\bar{\Delta}^n=\left\{ x \in [0,1]^n \subset \mathbb{R}^n \left| \sum_{i=1}^n x^i=1 \right\}. \right.
\end{equation*}
Let $\{x_{(k)} \in \Delta^n,k=1,..,n\}$ be $n$ fixed points in $\Delta^n$ with the property
\begin{equation}
\sum_{k=1}^n x_{(k)} = n e \label{E:Label08}
\end{equation}
where $e = (\frac{1}{n},...,\frac{1}{n}) \in \Delta^n$. These $n$ points $x_{(k)}$ will represent the outcomes of the weak measurements that we do, and we will refer to them as the fundamental steps for our discrete stochastic process. The measurement operators that we construct out of them are
\begin{equation}
\hat{N}_{(k)} = \sum_{i=1}^n \sqrt{x_{(k)}^i} \hat{P}_i \label{E:Label09}
\end{equation}
with the completeness condition satisfied:
\begin{align*}
& \sum_{k=1}^n \hat{N}_{(k)}^{\dagger} \hat{N}_{(k)} = \sum_{k,i,j=1}^n \sqrt{x_{(k)}^i} \hat{P}_i \sqrt{x_{(k)}^j} \hat{P}_j \\
& = \sum_{k,i=1}^n \sqrt{x_{(k)}^i x_{(k)}^j} \delta_{ij} \hat{P}_i = \sum_{k,i=1}^n x_{(k)}^i \hat{P}_i \\
& = n \sum_{i=1}^n e^i \hat{P}_i = \hat{I}.
\end{align*}
This means that the measurement operators \eqref{E:Label09} define a complete quantum measurement. If we measure the system using the above measurement operators over and over again, after a long enough time (strictly speaking in the limit of time going to infinity) the system is bound to end up in one of states \eqref{E:Label33} with the right probability. After every measurement we get a measurement outcome $(k), k \in \{1,...,n\}$ so after $s$ steps we have a sequence $(k_1, k_2,..., k_s)$ of outcomes. The state of the system at that time has changed to
\begin{align}
\ket{\psi_s} & = \ket{\psi_{(k_s,...,k_1)}} = \frac{1}{\sqrt{p_{k_s ... k_1}}} \hat{N}_{(k_s)} ... \hat{N}_{(k_1)} \ket{\psi_0} \notag \\
& = \frac{1}{\sqrt{p_{k_s ... k_1}}} \sum_{i=1}^n \sqrt{x_{(k_s)}^i ... x_{(k_1)}^i} \hat{P}_i \ket{\psi_0} \notag \\
& = \sum_{i=1}^n \sqrt{\frac{x_{(k_s)}^i ... x_{(k_1)}^i p_i^0}{p_{k_s ... k_1}}} \ket{\bar{\psi}_i} \label{E:Label01}
\end{align}
where
\begin{align*}
p_{k_s ... k_1} & = \bra{\psi_0} \hat{N}^{\dagger}_{(k_1)} ... \hat{N}^{\dagger}_{(k_s)} \hat{N}_{(k_s)} ... \hat{N}_{(k_1)} \ket{\psi_0} \\
& = \sum_{i=1}^n x_{(k_s)}^i ... x_{(k_1)}^i \bra{\psi_0} \hat{P_i} \ket{\psi_0} \\
& = \sum_{i=1}^n x_{(k_s)}^i ... x_{(k_1)}^i p_i^0.
\end{align*}
Equation (\ref{E:Label01}) can be rewritten as
\begin{align}
\ket{\psi_s} = \sum_{i=1}^n \sqrt{\tilde{x}_s^i} \ket{\bar{\psi}_i} \label{E:Label34}
\end{align}
with
\begin{align}
\tilde{x}_s^i = \frac{x_{(k_s)}^i ... x_{(k_1)}^i p_i^0}{p_{k_s ... k_1}}. \label{E:Label03}
\end{align}
Because
\begin{align*}
\sum_{i=1}^n \tilde{x}_s^i = \frac{\sum_{i=1}^n x_{(k_s)}^i ... x_{(k_1)}^i p_i^0}{p_{k_s ... k_1}} = 1
\end{align*}
it follows that $\tilde{x}_s \in \Delta^n$. In these new coordinates the initial state of the system is just $\ket{\psi_0} = \sum_{i=1}^n \sqrt{\tilde{x}_0^i} \ket{\bar{\psi}_i}$ where $\tilde{x}_0^i = p_i^0$.
Another way of writing equation (\ref{E:Label01}) is
\begin{align}
\ket{\psi_s} = \sum_{i=1}^n \sqrt{\frac{x_s^i p_i^0}{\sum_{l=1}^n x_s^l p_l^0}} \ket{\bar{\psi}_i} \label{E:Label02}
\end{align}
with
\begin{align}
x_s^i = \frac{x_{(k_s)}^i ... x_{(k_1)}^i}{\sum_{l=1}^n x_{(k_s)}^l ... x_{(k_1)}^l}. \label{E:Label04}
\end{align}
Here again $x_s \in \Delta^n$. Then $\ket{\psi_0}$ is given by (\ref{E:Label02}) with $x_0^i = \frac{1}{n}$.

Let us stress here that both formulas \eqref{E:Label34} and \eqref{E:Label02} imply that the state of the system changes after each weak measurement, generated by \eqref{E:Label09}, and that there is a correspondence between the evolving vector $\tilde{x}_s=(\tilde{x}^1_s,...,\tilde{x}^n_s)$ or $x_s=(x^1_s,...,x^n_s)$, the time evolution of each given by formula \eqref{E:Label03} and \eqref{E:Label04} respectively, and the evolving state of the system $\ket{\psi_s}$.

Another observation is that instead of keeping track of the sequence of outcomes $(k_1, k_2,...)$ we can equally well use the sequence of points $(\tilde{x}_1, \tilde{x}_2,...)$ or $(x_1, x_2,...)$ in $\Delta^n$. There is a one-to-one correspondence between sequences of outcomes and sequences of {\it permissible} points in $\Delta^n$. By a ``permissible point'' we mean a point for which there exists at least one sequence of outcomes $(k_1, k_2,..., k_s)$ that generates that point, by formula (\ref{E:Label03}) or (\ref{E:Label04}) respectively, given some initially chosen and fixed fundamental steps $\{x_{(k)}\}$. To prove that all three type of sequences carry the same information we use the fact that $\Delta^n$ can be given a group structure:
\begin{definition}
$\Delta^n$ is an abelian group with multiplication $\star: \Delta^n \times \Delta^n \rightarrow \Delta^n$ where $x \star y$ for $x, y \in \Delta^n$ is defined as
\begin{align}
(x \star y)^i = \frac{x^i y^i}{\sum_{k=1}^n x^k y^k}. \label{E:Label05}
\end{align}
\end{definition}
We also define the Hadamard product $x \circ y$ of two vectors $x$ and $y$ of the same dimension by
\begin{align*}
(x \circ y)^i = x^i y^i ,
\end{align*}
and the trace $Tr(x)$ of a vector $x$, which is defined as
\begin{align*}
Tr(x) = \sum_{i=1}^n x^i.
\end{align*}
Using these notations, equation (\ref{E:Label05}) reads as
\begin{align*}
x \star y = \frac{x \circ y}{Tr(x \circ y)}.
\end{align*}
The identity $e$ of the group $\Delta^n$ is $e = (1/n,...,1/n)$, and the inverse $x^{-1}$ of an element $x$ is $\frac{\bar{x}}{Tr(\bar{x})}$ where $\bar{x} = \left(\frac{1}{x^1},..., \frac{1}{x^n} \right)$.

Now it is easy to see that given a sequence of permissible points $(x_1, x_2,..., x_s)$, generated by the measurement procedure with some pre-fixed fundamental steps, every element of the sequence $(x_1, x_2 x_1^{-1},..., x_s x_{s-1}^{-1})$ is one of the fundamental steps $x_{(k)}$ for some $k$. Here $x^{-1}$ is the inverse of $x$ with respect to the group multiplication. So $(x_1, x_2 x_1^{-1},..., x_s x_{s-1}^{-1}) = (x_{(k_1)}, x_{(k_2)},..., x_{(k_s)})$ for some $(k_1,..., k_s)$ chosen appropriately, which is exactly the sequence of outcomes. It's as easy to prove the same for a sequence $(\tilde{x}_1, \tilde{x}_2,..., \tilde{x}_s)$.

The value of keeping track of sequences of permissible points is that we need only the last element $x_s$ of the sequence and the last outcome $i_{s+1}$ from the $(s+1)$-th measurement in order to find the state of the system $\ket{\psi_{s+1}}$ at time $s+1$: it is given by formula (\ref{E:Label02}) with
\begin{align}
x_{s+1} = x_s \star x_{(i_{s+1})} = \frac{x_s \circ x_{(i_{s+1})}}{Tr(x_s \circ x_{(i_{s+1})})}. \label{E:Label06}
\end{align}

We can make use of this result by making a correspondence between the measurement procedure and a classical discrete stochastic process. Let $(\Omega,\mathfrak{F},\mathbb{P})$ be a probability space. We define the stochastic process as $x: \mathbb{N}_0 \times \Omega \rightarrow P \Delta^n$ by giving its distribution law. Here $P \Delta^n$ is the set of permissible points for a concrete choice of fundamental steps ${x_{(k)}}$ and $\mathbb{N}_0 = \{0, 1, 2,...\}$ is the time set. Usually we will denote $x(s, \omega)$ as $x_s (\omega)$ for $\omega \in \Omega$. The {\textit{probability distribution} of $x_s$ thought of as random variable, is $d_s(\underline{x}) = \sum p_{k_1,..., k_s}$ where the sum is over all sequences of outcomes $(k_1,..., k_s)$ for which
\begin{align}
\underline{x} = \frac{x_{(k_s)}^i ... x_{(k_1)}^i}{\sum_{l=1}^n x_{(k_s)}^l ... x_{(k_1)}^l};
\end{align}
or in other words, for which the last entry in the sequence with $s$ elements is exactly $\underline{x}$.
If there is no such sequence with $s$ elements then $d_s(\underline{x})=0$. The reinterpretation of formula (\ref{E:Label06}) in these new terms is that the process $x$ is \textit{Markovian}.

\section{Continuous process for projective measurements}

We will now construct a \textit{continuous} quantum measurements process which decomposes the strong projective measurement (\ref{E:Label07}). Unlike the discrete case, we are now not restricted to performing the same measurement at every time step (in a sense, the fundamental steps can be chosen to be different). These measurements can be arbitrary, but they must depend smoothly on the state of the system at the current moment and not explicitly on the time.  (This will be vital when we take the next step to decomposing generalized measurements, where the measurements {\it must} be different at different steps.)

Let's make these ideas precise. As in the discrete case, we will concentrate on the evolution of the classical state analog of the system $X_t \in \Delta^n$ rather than its quantum state $\ket{\psi_t}$. The measurement on the system is now given by a Lesbegue-measurable set $D(x; \delta) \subset \Delta^n$, of vectors which is a neighborhood of $e \in \Delta^n$, and satisfies a condition analogous to (\ref{E:Label08}}):
\begin{align}
\int\limits_{D(x; \delta)} z dh_{\star}(z) = e \int\limits_{D(x; \delta)} dh_{\star}(z) = e A(x; \delta) \label{E:Label18}
\end{align}
with $A(x; \delta)$ being the $(n-1)$-dimensional volume of $D(x; \delta)$ and $h_{\star}(x)$ the Haar measure on $\Delta^n$. The other condition needed is
\begin{align}
\lim_{\delta \to 0} d(x; \delta) = \lim_{\delta \to 0} \sup_{z \in D(x; \delta)} d(z,e) = 0 \label{E:Label16}
\end{align}
where $d(z,x)$ is the standard distance function on $\Delta^n$ thought of as a subset of $\mathbb{R}^n$.
The set $D(x; \delta)$ depends on the current state of the system $x$ and on a small positive parameter $(0 \leq \delta \ll 1)$ which at some point we will let go to zero, thus getting a continuous stochastic process. The measurement operators are given as in (\ref{E:Label09}):
\begin{equation}
\hat{N} (z,x; \delta) = \left(\frac{n}{A}\right)^{1/2} \sum_{i=1}^n \sqrt{z^i} \hat{P}_i \text{ with } z \in D(x; \delta) \label{E:Label10}
\end{equation}
where the completeness condition is given by
\begin{align*}
& \int\limits_{D(x; \delta)} \hat{N}^{\dagger} (z,x) \hat{N} (z,x) dh_{\star}(z) \\
& = \frac{n}{A} \sum_{i,j=1}^n \hat{P}_i \hat{P}_j \int\limits_{D(x; \delta)} \sqrt{z^i} \sqrt{z^j} dh_{\star}(z) \\
& = \frac{n}{A} \sum_{i=1}^n \hat{P}_i \int\limits_{D(x; \delta)} z^i dh_{\star}(z) \\
& = \frac{n}{A} \sum_{i=1}^n \hat{P}_i e^i A = \hat{I}
\end{align*}
When we perform a measurement we get an outcome $z^o$ from the set $D(x; \delta)$. The new state of the system $z$ is given by formula \eqref{E:Label06}:
\begin{align*}
z = x \star z^o = \frac{x \circ z^o}{Tr(x \circ z^o)}.
\end{align*}
The probability density for the system to be in state $z \in D_x^{\star}(x; \delta)$ is
\begin{align}
\mu(z) = \bra{\psi_t} \hat{N}^{\dagger} (z^o,x; \delta) \hat{N} (z^o,x; \delta) \ket{\psi_t} \label{E:Label13}
\end{align}
where $\ket{\psi_t}$ is given by \eqref{E:Label02} and $D_u^{\star}(x; \delta)$ is the left translation of $D(x; \delta)$ by $u$ (sometimes the notation $u \star D(x; \delta)$ could also be used) defined as
\begin{align*}
D_u^{\star}(x; \delta) = \{y \in \Delta^n | y = u \star y^o, \text{ for some } y^o \in  D(x; \delta) \}.
\end{align*}
Expanding \eqref{E:Label13} we get
\begin{align}
\mu(z) & = \frac{n}{A} \sum_{k,l=1}^n \sqrt{\frac{x^k p_k^0}{\sum_{m=1}^n x^m p_m^0}} \sqrt{\frac{x^l p_l^0}{\sum_{m=1}^n x^m p_m^0}} \times \notag \\
& \times \sum_{i,j=1}^n \sqrt{(z^o)^i} \sqrt{(z^o)^j} \bra{\bar{\psi}_k} \hat{P}_i \hat{P}_j \ket{\bar{\psi}_l} \notag \\
& = \frac{n}{A} \sum_{i=1}^n \frac{x^i p_i^0 (z^o)^i}{\sum_{m=1}^n x^m p_m^0} = \frac{n}{A} \frac{Tr(x \circ p^0 \circ z^o)}{Tr(x \circ p^0)} \notag \\
& = \frac{n}{A} \frac{Tr(x \circ p^0 \circ (x^{-1} \star z))}{Tr(x \circ p^0)} \label{E:Label11}
\end{align}
and if $z \in \Delta^n \backslash D_x^{\star}(x; \delta)$ then $\mu(z) = 0$.

We want to emphasize that performing a measurement with operators \eqref{E:Label10} on the system at time $t$ gives the new state of the system at a later time $t + \varepsilon$. For the stochastic process to be continuous we want that $\varepsilon \to 0$ as $\delta \to 0$.

The easiest way to derive the stochastic differential equation for the process $X$ requires the concept of \textit{stopping time}. To give a proper definition we need to consider not just a probability space but a filtered probability space. More information on probability spaces, stochastic processes, stochastic differential equations and It\"{o}'s calculus can be found in \cite{Ba,Kan,Kar,Ev,IkWa}.
\begin{definition}
Let $(\Omega,\mathfrak{F},\mathbb{P})$ be a probability space. A family $\{\mathfrak{F}(t)\}_{t \geq 0}$ of $\sigma$-algebras $\mathfrak{F}(t) \subseteq \mathfrak{F}$ is \emph{a filtration} on the probability space $(\Omega,\mathfrak{F},\mathbb{P})$ if $\mathfrak{F}(s) \subseteq \mathfrak{F}(t)$ for $s \leq t$. We are going to call $(\Omega,\mathfrak{F}(t)\}_{t \geq 0},\mathfrak{F},\mathbb{P})$ \emph{a filtered probability space}.
\end{definition}
In this paper we will always assume that the filtered probability space satisfies all the needed hypotheses. $\mathfrak{F}$ is $\mathbb{P}$-complete, $\mathfrak{F}_t$ contains all the null sets in $\mathfrak{F}$, $\mathfrak{F}_t$ is right-continuous. Also, all the stochastic processes are to be adapted to the filtration.
\begin{definition}
A process $X:\mathbb{R}_+ \times \Omega \rightarrow V$ (with $(V, \mathfrak{V})$ a measurable space) is \emph{adapted} if $X_t = X(t) := X(t,\cdot)$ is $\mathfrak{F}_t$-measurable for all $t \in \mathbb{R}_+$.
\end{definition}
\begin{definition}
A process $X:\mathbb{R}_+ \times \Omega \rightarrow V$ (with $(V, \tau)$ a topological space) is \emph{continuous} if $X(t, \omega)$ is a continuous function of $t$ for all $\omega \in \Omega$.
\end{definition}
Now we can give the definition of a stopping time:
\begin{definition}
A random variable $\tau:\Omega \rightarrow [0,\infty]$ is called \emph{a stopping time} with respect to $\{\mathfrak{F}_t\}_{t \geq 0}$ provided
\begin{align*}
\{\tau \leq t\} = \tau^{-1}([0,t]) \in \mathfrak{F}_t \text{ for all } t \geq 0.
\end{align*}
\end{definition}
\begin{definition}
Given a stochastic process $X:\mathbb{R}_+ \times \Omega \rightarrow V$ and a stopping time $\tau$ the stopped process $X^{\tau}$ is defined as
\begin{align*}
X_t^{\tau} = X_{\min \{t,\tau \}}
\end{align*}
\end{definition}
The stopping times that we will deal with are called hitting times.
\begin{lemma}
Let $X:\mathbb{R}_+ \times \Omega \rightarrow V$ be an adapted stochastic process and $S \subset V$ a measurable set. Then
\begin{align*}
\tau := \inf\{t \geq 0 \left| X_t \in S \} \right.
\end{align*}
is a stopping time.
\end{lemma}
The hitting time is just the first time when the stochastic process touches the set $S$.

Now we can derive the stochastic differential equation for our stochastic process $X_t$. Let $S \in \Delta^n$ and $B \subset \partial S$ be a part of the boundary $\partial S$ of $S$. Let $P_B^{X}(X_0)$ be the probability that, when the stochastic process $X$ with time evolution generated by the measurement operators (\ref{E:Label10}) and initial condition $X_0$ hits the boundary, $\partial S$, it hits it in the subset $B$ rather than in $\partial S \backslash B$. Let's denote by $\tau_{\partial S}$ the hitting time of the process of the the boundary $\partial S$.

What we want is to find a second-order differential equation for $P_B^{X}(x)$. Let's assume that at time $t$ the system state is localized at the point $x$; $X_t = x$. Performing the measurement with operators \eqref{E:Label10} gives us the state of the system $X_{t + \varepsilon}$ at time $t + \varepsilon$ which is a random variable with probability density distribution given by \eqref{E:Label11}. Now we need the following simple theorem:
\begin{theorem}
Let $X_t$ be the value of the stochastic process defined above at some time ${t < \tau_{\partial S}}$. Then if $X^{\to}$ is the stochastic process with time evolution generated by the measurement operators (\ref{E:Label10}) but with initial condition $X_0^{\to} = X_t$ we have
\begin{equation*}
P_B^{X}(X_0) = P_B^{X^{\to}}(X_t).
\end{equation*}
\end{theorem}
The statement of the theorem is trivial because $X_t^{\to} = X_{t+T}$.

From the theorem follows that
\begin{equation}
P_B^{X}(x) = P_B^{X}(X_{t + \varepsilon}). \label{E:Label12}
\end{equation}
$P_B^{X}(X_0)$ is a conditional probability---it's the probability for hitting $B$ given that the initial condition is $X_0$, and thus it satisfies the following property
\begin{align}
P_B^{X}(X_0) & = \int\limits_{\Delta^n} P_B^{X}(z) \mathbb{P}(X_0^{-1}(dh_{\star}(z))) \notag \\
& = \int\limits_{\Delta^n} P_B^{X}(z) \Theta_{X_0}(dh_{\star}(z)) \notag \\
& = \int\limits_{\Delta^n} P_B^{X}(z) \theta_{X_0}(z) dh_{\star}(z)
\end{align}
where $\Theta_{X_0}(z) = \mathbb{P}(X_0^{-1}(z))$ is the probability distribution of $X_0$ and $\theta_{X_0}(z)$ the probability density with respect to the Haar measure $h_{\star}$. Substituting this in \eqref{E:Label12} we have
\begin{align}
P_B^{X}(x) = P_B^{X}(X_{t + \varepsilon}) = \int\limits_{D_x^{\star}(x; \delta)} P_B^{X}(z) \theta_{X_{t + \varepsilon}}(z) dh_{\star}(z) \label{E:Label14}
\end{align}
with $\theta_{X_{t + \varepsilon}}(z)$ given by \eqref{E:Label11}
\begin{align*}
\theta_{X_{t + \varepsilon}}(z) = \frac{n}{A} \frac{Tr(x \circ p^0 \circ (x^{-1} \star z))}{Tr(x \circ p^0)}.
\end{align*}
After changing the coordinates $z \rightarrow x \star z$, given that the Haar measure is invariant under left translations ($h(A) = h(x \star A)$ for $\forall x \in \Delta^n$ and $A \subset \Delta^n$), \eqref{E:Label14} becomes
\begin{align}
P_B^{X}(x) = \int\limits_{D(x; \delta)} P_B^{X}(x \star z) \tth_{X_{t + \varepsilon}}(z) dh_{\star}(z) \label{E:Label15}
\end{align}
with
\begin{align}
\tth_{X_{t + \varepsilon}}(z) = \frac{n}{A} \frac{Tr(x \circ p^0 \circ z)}{Tr(x \circ p^0)}. \label{E:Label20}
\end{align}
Equations \eqref{E:Label15} and \eqref{E:Label20} are the starting point for deriving the differential equation for $P_B^{X}(x)$. We first do an integral estimate that is needed to take the limit $\delta \to 0$ in \eqref{E:Label15}. We want to estimate the integral
\begin{align*}
Z_k^{i_1,...,i_k} (x; \delta) & = \int\limits_{D(x; \delta)} ((z - e)^{\otimes k})^{i_1,...,i_k} d h_{\star}(z) \\
& = \int\limits_{D(x; \delta)} (z - e)^{i_1}...(z - e)^{i_k} d h_{\star}(z).
\end{align*}
Using the notation in \eqref{E:Label16} for the $i_k$-th component of $z-e$ thought of as a vector in $\mathbb{R}^n$ we get
\begin{align*}
(z - e)^{i_k} \leq d(x; \delta) \text{ for } z \in D(x; \delta)
\end{align*}
and it follows immediately that
\begin{align*}
(z - e)^{i_1}...(z - e)^{i_k} \in \mathcal{O}(d^k(\delta)) \text{ as } \delta \to 0.
\end{align*}
In the end
\begin{align*}
Z&_k^{i_1,...,i_k} (x; \delta) \leq d^k(x; \delta) \int\limits_{D(x; \delta)} dh_{\star}(z) \\
& = d^k(x; \delta) A(x; \delta)  \in \mathcal{O}(d^k(\delta) A(\delta)) \text{ as } \delta \to 0.
\end{align*}
From this it is a trivial consequence that
\begin{align}
\frac{Z_k^{i_1,...,i_k} (x; \delta)}{d^2(\delta) A(\delta)} \in \mathcal{O}(d^{k-2}(\delta)) \text{ as } \delta \to 0. \label{E:Label17}
\end{align}
Now we can take the limit $\delta \to 0$ of \eqref{E:Label15}, dividing it first on both sides by $d^2(\delta)$. For this purpose we first expand the integrand in a Taylor series. Because of \eqref{E:Label17}, all terms of order three or more in the series go to zero as $\delta \to 0$, and so we need to keep only terms of order zero, one and two. To simplify the notation we denote $p(z) = P_B^{X}(z)$, $\tth(z) = \tth_{X_{t + \varepsilon}}(z)$ and $y = x \star z$. We have
\begin{align*}
p&(x) = p(y)|_{z=e} + \left. \frac{\partial p(y)}{\partial z} \right|_{z=e} \cdot (z-e) \\
& + \frac{1}{2} (z-e) \cdot \left. \frac{\partial^2 p(y)}{\partial z^2} \right|_{z=e} \cdot (z-e) + \mathcal{O}((z-e)^3) \\
& = p(x) + \left. \frac{\partial p(y)}{\partial y} \right|_{y=x} \cdot \left. \frac{\partial y}{\partial z} \right|_{z=e} \cdot (z-e) \\
& + \frac{1}{2} (z-e) \cdot \left. \frac{\partial y}{\partial z} \right|_{z=e} \cdot \left. \frac{\partial^2 p(y)}{\partial y^2} \right|_{y=x} \cdot \left. \frac{\partial y}{\partial z} \right|_{z=e} \cdot (z-e) \\
& + \frac{1}{2} \left. \frac{\partial p(y)}{\partial y} \right|_{y=x} \cdot \left((z-e) \cdot \left. \frac{\partial^2 y}{\partial z^2} \right|_{z=e} \cdot (z-e) \right) \\
& + \mathcal{O}((z-e)^3).
\end{align*}
Above $\cdot$ denotes the usual dot product in $\mathbb{R}^n$. Substituting this in \eqref{E:Label15} while recalling that
\begin{align*}
\int\limits_{D(x; \delta)} \tth_{X_{t + \varepsilon}}(z) dh_{\star}(z) = 1,
\end{align*}
we get
\begin{align}
p(x) & = \left( p(y) + \frac{\partial p(y)}{\partial y} \cdot \frac{\partial y}{\partial z} \cdot \mathfrak{T}_1 (x; \delta) \right. \notag \\
& + \frac{1}{2} Tr \left( \frac{\partial y}{\partial z} \cdot \frac{\partial^2 p(y)}{\partial y^2} \cdot \frac{\partial y}{\partial z} \cdot \mathfrak{T}_2 (x; \delta) \right) \notag \\
& \left. \left. + \frac{1}{2} \frac{\partial p(y)}{\partial y} \cdot Tr \left( \frac{\partial^2 y}{\partial z^2} \cdot \mathfrak{T}_2 (x; \delta) \right) \right) \right|_{y=x,z=e} \notag \\
& + \mathcal{O}(d^3(\delta)A(\delta)), \label{E:Label30}
\end{align}
where
\begin{align}
\mathfrak{T}_k (x; \delta) = \int\limits_{D(x; \delta)} (z-e)^{\otimes k} \tth(z) dh_{\star}(z). \label{E:Label29}
\end{align}
Expanding $\tth(z)$ in a Taylor series,
\begin{align*}
\tth(z) & = \tth(e) + \left. \frac{\partial \tth(z)}{\partial z} \right|_{z=e} \cdot (z-e) \notag \\
& = \frac{1}{A} + \frac{n}{A} (x \star p^0) \cdot (z - e),
\end{align*}
and substituting the series into \eqref{E:Label29}, we get
\begin{align}
\mathfrak{T}_k (x; \delta) = \frac{1}{A} Z_k (x; \delta) + \frac{n}{A} (x \star p^0) \cdot Z_{k+1} (x; \delta).
\end{align}
Now we can simplify \eqref{E:Label30}, and after dividing both sides of the equation by $d^2(\delta)$ we get
\begin{align}
0 & = \left. \frac{\partial p(y)}{\partial y} \right|_{y=x} \cdot \left. \frac{\partial y}{\partial z} \right|_{z=e} \cdot \frac{Z_1 (x; \delta)}{d^2(\delta)A(\delta)} \notag \\
& + \left. \frac{\partial p(y)}{\partial y} \right|_{y=x} \cdot \left. \frac{\partial y}{\partial z} \right|_{z=e} \cdot \frac{n Z_2 (x; \delta)}{d^2(\delta)A(\delta)} \cdot (x \star p^0) \notag \\
& + \frac{1}{2} Tr \left( \left. \frac{\partial y}{\partial z} \right|_{z=e} \cdot \left. \frac{\partial^2 p(y)}{\partial y^2} \right|_{y=x} \cdot \left. \frac{\partial y}{\partial z} \right|_{z=e} \cdot \frac{Z_2 (x; \delta)}{d^2(\delta)A(\delta)} \right) \notag \\
& + \frac{1}{2} \left. \frac{\partial p(y)}{\partial y} \right|_{y=x} \cdot Tr \left( \left. \frac{\partial^2 y}{\partial z^2} \right|_{z=e} \cdot \frac{Z_2 (x; \delta)}{d^2(\delta)A(\delta)} \right) \notag \\
& + \mathcal{O}(d(\delta)A(\delta)). \label{E:Label19}
\end{align}
The term in the first line of \eqref{E:Label19} is zero because \eqref{E:Label18} can be rewritten as
\begin{align*}
\int\limits_{D(x; \delta)} (z - e) dh_{\star}(z) = 0 = Z_1 (x; \delta).
\end{align*}
Now we can take the limit $\delta \to 0$. As
\begin{align*}
\frac{Z_2 (x; \delta)}{d^2(\delta)A(\delta)} \in \mathcal{O}(1) \text{ as } \delta \to 0
\end{align*}
it has a finite limit which we denote by $\eta(x)$. From its definition it follows that this matrix is positive with at least one eigenvalue equal to zero, because
\begin{align}
& \sum_{i=1}^n \int\limits_{D(x; \delta)} (z-e)^i (z-e)^k dh_{\star}(z) \notag \\
& = \int\limits_{D(x; \delta)} \sum_{i=1}^k (z^i-e^i) (z-e)^k dh_{\star}(z) = 0. \label{E:Label27}
\end{align}
We will assume that all other eigenvalues of $\eta(x)$ are different from zero. Because $\mathcal{O}(d(\delta)A(\delta)) \to 0$ as $\delta \to 0$, we finally derive the second-order differential equation for $p(x)$:
\begin{align}
&n \frac{\partial p}{\partial x} \cdot \left. \frac{\partial y}{\partial z} \right|_{z=e} \cdot \eta(x) \cdot (x \star p^0) \notag \\
& + \frac{1}{2} Tr \left( \left. \frac{\partial y}{\partial z} \right|_{z=e} \cdot \frac{\partial^2 p}{\partial x^2} \cdot \left. \frac{\partial y}{\partial z} \right|_{z=e} \cdot \eta(x) \right) \notag \\
& + \frac{1}{2} \frac{\partial p}{\partial x} \cdot Tr \left( \left. \frac{\partial^2 y}{\partial z^2} \right|_{z=e} \cdot \eta(x) \right) = 0. \label{E:Label22}
\end{align}
Differentiating $y = x \star z$ we get
\begin{align}
\left. \frac{\partial y^i}{\partial z^j} \right|_{z=e} = n x^i (\delta_j^i - x^j) \label{E:Label21}
\end{align}
and
\begin{align}
\left. \frac{\partial^2 y^i}{\partial z^j \partial z^k} \right|_{z=e} = n^2 x^i (2 x^j x^k - x^j \delta_k^i - x^k \delta_j^i).
\end{align}
The matrix \eqref{E:Label21} has a pseudo-inverse given by $\frac{1}{n x^i}(\delta_i^k - \frac{1}{n})$ and so it satisfies
\begin{align*}
\sum_{i=1}^n \left. \frac{\partial y^i}{\partial z^j} \right|_{z=e} \frac{1}{n x^i}(\delta_i^k - \frac{1}{n}) & = \sum_{i=1}^n (\delta_j^i - x^j) \left(\delta_i^k - \frac{1}{n}\right) \notag \\
& = \delta_j^k - \frac{1}{n}.
\end{align*}
Substituting these in \eqref{E:Label22} we get the final simple form of our equation
\begin{align}
\frac{1}{2} g^{ij}(x) \frac{\partial^2 p}{\partial x^i \partial x^j} + b_i(x) g^{ij}(x) \frac{\partial p}{\partial x^j} = 0
\end{align}
where
\begin{align}
g^{ij}(x) = \frac{1}{n^2} \left. \frac{\partial y^i}{\partial z^{\alpha}} \right|_{z=e} \eta^{\alpha \beta}(x) \left. \frac{\partial y^j}{\partial z^{\beta}} \right|_{z=e} \label{E:Label23}
\end{align}
and
\begin{align}
b_j(x) = \frac{(x \star p^0)^j}{x^j} = \frac{p_j^0}{Tr(x \circ p^0)}.
\end{align}

It's simple to derive the stochastic equations for the process $X_t$ having equation \eqref{E:Label23} for the probability $P_B^X$. We use the following theorems to that purpose
\begin{theorem}
(It\"{o}'s formula for stopping times) Given a stochastic process satisfying $dX_t^i = \gamma^i (X, t) dt + \sum_{j=1}^n \sigma_j^i(X, t) dW^j$, a stopping time $\tau$ and a $C^{2,1}$-function $u(x,t)$, it follows that
\begin{align}
\left. u(X_t, t) \right|_{0}^{\tau} = \int\limits_0^\tau \left. \left( \frac{\partial u}{\partial t} + \mathcal{L}u \right) \right|_{t=s} ds + \int\limits_0^{\tau} \partial u \cdot \sigma \cdot dW \label{E:Label24}
\end{align}
where
\begin{align}
\mathcal{L}u = \frac{1}{2} \sum_{i,j=1}^n \Sigma^{ij} \frac{\partial^2 u}{\partial x^i \partial x^j} + \sum_{i=1}^n \gamma^i \frac{\partial u}{\partial x^i} \text{, } \Sigma^{ij} = \sum_{k=1}^n \sigma_k^i \sigma_k^j.
\end{align}
\end{theorem}
Taking expectations of \eqref{E:Label24} we get
\begin{align}
\mathbb{E} (u(X_{\tau}, \tau)) - \mathbb{E} (u(X_0, 0)) = \mathbb{E} \left( \int\limits_0^\tau \left. \left( \frac{\partial u}{\partial t} + \mathcal{L}u \right) \right|_{t=s} ds \right).
\end{align}
\begin{theorem}
Given the conditions in the theorem above, the $C^2$-function $u(x)$ defined on some smooth, bounded domain $U$ given by
\begin{align*}
u(x) = \mathbb{E} (h(X_{\tau})) \text{ with } X_0 =
\end{align*}
satisfies the boundary value problem
\begin{align*}
\mathcal{L}u = 0 \text{ in } U, \\
u = h \text{ on } \partial U.
\end{align*}
\end{theorem}
We can now apply this to our case:
\begin{align}
\mathcal{L} P_B^X = 0 & \text{ in } S, \notag \\
P_B^X = 1 & \text{ on } \partial B, \notag \\
P_B^X = 0 & \text{ on } \partial S \backslash B \label{E:Label25}
\end{align}
with
\begin{align}
\mathcal{L}u = \frac{1}{2} g^{ij}(x) \frac{\partial^2 u}{\partial x^i \partial x^j} + b_i(x) g^{ij}(x) \frac{\partial u}{\partial x^j}. \label{E:Label32}
\end{align}
From these theorems we see that if we require that the stochastic process satisfies the following stochastic differential equation it will satisfy equations \eqref{E:Label25} and \eqref{E:Label32}, too
\begin{align}
dx_t^i = \sum_{j=1}^n g^{ij}(x_t) b_j (x_t) dt + \sum_{k=1}^n a_k^i (x_t) dW^k \label{E:Label26}
\end{align}
where $a(x)$ is the unique square root of $g(x)$:
\begin{align}
g^{ij} = \sum_{k=1}^n a_k^i a_k^j.
\end{align}
\eqref{E:Label26} is the equation we consider for governing the evolution of the continuous measurement process. We want to point out the form of the equation is not accidental. Such equation arise when one considers stochastic processes in local coordinates on Riemannian manifolds equipped with an adapted (but not torsion-free) connection. When one considers Brownian motion on manifolds the same equation appears but the term in front of $dt$ involves the Levy-Civita connection associated with the metric $g$. For further reading on stochastic processes on manifolds one can turn to references \cite{Hs,El,Em,Ra,La}.

Now we prove that if the metric $g(x)$ is chosen properly, the stochastic process reproduces the strong measurement at long times.
\begin{theorem}
Let the metric $g$ be invariant under the action of $\Delta^n$ on itself and of the symmetric group $S_n$. Then the stochastic process satisfying \eqref{E:Label26} starting from $X_0 = e$ ends in one of the vertices $v_{(k)}$ of $\bar{\Delta}^n$ with probability $p_k^0$ in the limit $t \to \infty$.
\end{theorem}
The action of $S_n$ on $\bar{\Delta}^n$ is defined as follows --- for $\sigma \in S_n$ and $x \in \bar{\Delta}^n$,
\begin{align*}
(\sigma x)^i = x^{\sigma(i)}.
\end{align*}
The components of the vertices $v_{(k)}^i$ are equal to $\delta_k^i$.
\newline \textit{Proof.} The conditions for invariance of the metric $g$ together with condition \eqref{E:Label27} fix the form of the metric very stringently --- the metric is unique up to a constant conformal factor and has the following form:
\begin{align}
g^{ij} (x) = \sum_{\alpha,\beta=1}^n x^i (\delta_{\alpha}^i - x^{\alpha}) \eta^{\alpha \beta} x^j (\delta_{\beta}^j - x^{\beta}),
\end{align}
where the matrix $\eta^{\alpha \beta}$ is constant and equal to the projector from $\mathbb{R}^n$ to $\Delta^n$:
\begin{align}
\eta^{\alpha \beta} = \delta_{\beta}^{\alpha} - \frac{1}{n}.
\end{align}
The stochastic equation \eqref{E:Label26} will have the form
\begin{align}
dx^i & = \sum_{\alpha,\beta=1}^n x^i (\delta_{\alpha}^i - x^{\alpha}) \eta^{\alpha \beta} x^j (\delta_{\beta}^j - x^{\beta}) b_j (x) dt \notag \\
& \qquad + \sum_{\alpha=1}^n x^i (\delta_{\alpha}^i - x^{\alpha}) \sigma_k^{\alpha} dW^k,
\end{align}
with $\sigma$ is the unique square root of $\eta$:
\begin{align}
\sigma_k^{\alpha} = \delta_k^{\alpha} -\frac{1}{n}.
\end{align}
If we change the coordinates from $x$ to $\tilde{x}$, defined by
\begin{align}
\tilde{x} = x \star p^0, \label{E:Label28}
\end{align}
the equation in these new coordinates looks simple:
\begin{align}
d\tilde{x}^i = \sum_{\alpha=1}^n \tilde{x}^i (\delta_{\alpha}^i - \tilde{x}^{\alpha}) \sigma_k^{\alpha} dW^k = \tilde{x}^i \sum_{k=1}^n (\delta_k^i - \tilde{x}^k) dW^k.
\end{align}
From this it immediately follows that the stochastic process $X_t$ in these coordinates is a local martingale; and as it takes values in the bounded set $\Delta^n$, it is a martingale. As the process is a $L_p$-integrable martingale for any $p \geq 1$, it follows by the martingale convergence theorem that there exists a random variable $\tilde{x}_{\infty}$ which is the limit of $\tilde{x}_t$ as $t \to \infty$. Now we will prove that the limit $\tilde{x}_{\infty}$ is localized on the vertices of $\Delta^n$. Let's denote the $m$-th moment of $\tilde{x}$ $\mathfrak{m}_m (\tilde{x}) = \sum_{i=1}^n (\tilde{x}^i)^m$. Then using It\"{o}'s calculus we get
\begin{align*}
d\mathfrak{m}_2 (\tilde{x}) & = \sum_{i,k=1}^n \tilde{x}^i (\delta_k^i - \tilde{x}^k) \tilde{x}^i (\delta_k^i - \tilde{x}^k) dt \\
& \qquad + 2 \sum_{i,k=1}^n (\tilde{x}^i)^2 (\delta_k^i - \tilde{x}^k) dW^k \\
& = \left( \mathfrak{m}_2 (\tilde{x}) - 2 \mathfrak{m}_3 (\tilde{x}) + \mathfrak{m}_2^2 (\tilde{x}) \right) dt \\
& \qquad + 2 \sum_{i,k=1}^n (\tilde{x}^i)^2 (\delta_k^i - \tilde{x}^k) dW^k.
\end{align*}
Taking the expectation value, we arrive at an ordinary differential equation for $\mathbb{E} (\mathfrak{m}_2)$:
\begin{align}
\frac{d\mathbb{E}(\mathfrak{m}_2)}{dt} = \mathbb{E} (\mathfrak{m}_2) - 2 \mathbb{E} (\mathfrak{m}_3) + \mathbb{E} \left( \mathfrak{m}_2^2 \right).
\end{align}
As the limit of $\tilde{x}_t$ as $t \to \infty$ exists, so does the limit of $\mathbb{E}(\mathfrak{m}_2 (\tilde{x}_t))$. This means that the time derivative of $\mathbb{E}(\mathfrak{m}_2 (\tilde{x}_t))$ goes to zero as time goes to infinity:
\begin{align}
\mathbb{E} (\mathfrak{m}_2 (\tilde{x}_{\infty})) - 2 \mathbb{E} (\mathfrak{m}_3 (\tilde{x}_{\infty})) + \mathbb{E} \left( \mathfrak{m}_2^2 \right (\tilde{x}_{\infty})) = 0.
\end{align}
This last equation implies that the range of the random variable $\tilde{x}_{\infty}$ is a subset of the set of all zeros of the function $\left( \mathfrak{m}_2 - 2 \mathfrak{m}_3 + \mathfrak{m}_2^2 \right) (x)$. It is easy to show that the roots of this function are exactly the vertices of $\bar{\Delta}^n$ by considering the inequality
\begin{align}
& \left( \mathfrak{m}_2 - 2 \mathfrak{m}_3 + \mathfrak{m}_2^2 \right) (x) = \sum_{i=1}^n \left[ (\tilde{x}^i)^2 \left( 1 - 2 \tilde{x}^i + \sum_{k=1}^n (\tilde{x}^k)^2 \right) \right] \notag \\
& \geq \sum_{i=1}^n (\tilde{x}^i)^2 (1 - 2 \tilde{x}^i + (\tilde{x}^i)^2) = \sum_{i=1}^n (\tilde{x}^i)^2 (1 - \tilde{x}^i)^2 \geq 0
\end{align}
with equality when $x \in \{ v_{(1)},..., v_{(n)} \}$. This proves that the stochastic process ends at one of the vertices of $\bar{\Delta}^n$ when time goes to infinity.

Let's denote the probability distribution function of $\tilde{x}_{\infty}$ by $\mathfrak{M} (x)$. Then
\begin{align}
\mathfrak{M} (x) =
\begin{cases}
q_k, &\text{for } x = v_{(k)}\\
0, &\text{for } x \neq v_{(k)} \text{ for } \forall k=1,...,n
\end{cases}
\end{align}
for some $q = (q_1,...,q_n) \in \Delta^n$.
As $\tilde{x}_t$ is a martingale it follows that
\begin{align}
\mathbb{E} (\tilde{x}_0) = \mathbb{E} (\tilde{x}_t) = \mathbb{E} (\tilde{x}_{\infty}).
\end{align}
In the $x$-coordinates the process starts from $x_0 = e$. By \eqref{E:Label28} the initial condition in the $\tilde{x}$-coordinates is $\tilde{x}_0 = p^0$. Then
\begin{align}
p_i^0 = \mathbb{E} (\tilde{x}_0^i) = \mathbb{E} (\tilde{x}_{\infty}^i) = \sum_{k=1}^n q_k v_{(k)}^i = \sum_{k=1}^n q_k \delta_k^i = q_i.
\end{align}
This proves the theorem.

Now we can describe the evolution of the state of the system when it is subjected to our feedback control scheme. The equations are a special case of equations \eqref{E:Label35} with $M_j = P_j$:
\begin{align}
d\ket{\psi_t} & = - \frac{1}{8} \sum_{j=1}^n (\hat{P}_j - \opav{\hat{P}_j}) (\hat{P}_j - \opav{\hat{P}_j}) \ket{\psi_t} dt \notag \\
& \qquad + \frac{1}{2} \sum_{i=1}^n (\hat{P}_i - \opav{\hat{P}_i}) \ket{\psi_t} dW^{i}, \label{E:Label36} \\
dx^i & =  g^{ij} \opav{\hat{P}_j} dt + a_{\alpha}^i dW^{\alpha},
\end{align}
with $\opav{\hat{P_j}} = \bra{\psi_t} \hat{P_j} \ket{\psi_t}$.
The notable thing in these equations, that distinguishes them from the equations describing the the decomposition of a generalized measurement, is that feedback is not needed. The state of the system $\ket{\psi_t}$ evolves in accordance with equation \eqref{E:Label36} which has no dependance on the vector $x$. As we will see below, that is not true for generalized measurement. In that case the equations for $\ket{\psi_t}$ and $x$ are coupled in a nontrivial way, and feedback is necessary.

\section{Continuous processes for generalized measurements}

Building from the above process on the classical state space $\Delta^n$, it is now easy to prove the existence of a stochastic process on the quantum state space that decomposes {\it any} kind of generalized measurement, given by measurement operators $\hat{M}_j \; (j=1,...,n)$ satisfying the usual completeness relation $\sum_{j=1}^n \hat{M}_j^{\dagger} \hat{M}_j = \hat{I}$, into continuous measurements.  (It is not necessary that these measurement operators commute.) To that purpose we will use the fact that the homomorphisms of stochastic differential equations(the transformations that preserve the structure of the equations) must be at least twice differentiable maps. We will give an example of a map $\mathcal{M}$ taking points in $\Delta^n$ and mapping them to operators acting on the state space of our quantum system. This map should satisfy the following three requirements: it should be twice differentiable in $\Delta^n$, equal to the identity $\hat{I}$ at the identity $e$ of $\Delta^n$, and equal to the measurement operators at the vertices of $\bar{\Delta}^n$.

Let $\hat{M}_j = \hat{U}_j \hat{L}_j$ be the left polar decomposition of $\hat{M}_j$, so $\hat{L}_j$ are positive and $\hat{U}_j$ are unitary. The map $\mathcal{\hat{M}}$ will be the product of two maps $\hat{\Upsilon} (x)$ and $\hat{\Lambda} (x)$: the first involving the unitaries $\hat{U}_i$, and the second the positive operators $\hat{L}_j$. If the hamiltonians corresponding to the unitaries $\hat{U}_j$ are $\hat{H}_j$, then we can choose the map $\hat{\Upsilon} (x)$ to be
\begin{align}
\hat{\Upsilon} (x) = exp \left( \frac{in}{n-1} \sum_{j=1}^n x^j \left( x^j - \frac{1}{n} \right) \hat{H}_j \right).
\end{align}
We can readily see that, as constructed, the map has all the required properties, and also has the property that $\hat{\Upsilon} (x)$ is unitary for all $x \in \Delta^n$. $\hat{\Lambda} (x)$ is given by
\begin{align}
\hat{\Lambda} (x) = f^{\frac{1}{2}} (x) \left( \sum_{j=1}^n x^j \hat{L}_j^{\dagger} \hat{L}_j \right)^{\frac{1}{2}}.
\end{align}
with $f(x) = 1 + n \sum_{j=1}^n x^j (1 - x^j)$. The square root above is well-defined because the the expression in parentheses is a positive operator for every $x \in \Delta^n$. We also note that $\hat{\Lambda} (x)$ is invertible for every $x \in \Delta^n$. This follows from the completeness property satisfied by the $\{\hat{L}_j\}$: $\sum_{j=1}^n \hat{L}_j^{\dagger} \hat{L}_j = \hat{I}$. The map $\mathcal{\hat{M}} (x)$ is given by $\hat{\Upsilon} (x) \hat{\Lambda} (x)$.

The quantum state $\rho$ will respectively evolve according to the following formula
\begin{align}
\rho_t = \frac{\mathcal{\hat{M}} (x_t) \rho_0 \mathcal{\hat{M}}^{\dagger} (x_t)}{Tr(\mathcal{\hat{M}}^{\dagger} (x_t) \mathcal{\hat{M}} (x_t) \rho_0)}. \label{E:Label31}
\end{align}

It is obvious that the constructed map is far from unique---there are infinitely many maps that satisfy the three required conditions. This shows that there are numerous ways to decompose a strong generalized measurement into weak measurements of the type that we are considering.

In the very special case where the measurement operators are {\it positive} and {\it commuting}, we will now show that the state evolves according to a generalized quantum state diffusion equation.  As the measurement operators are positive, the unitaries in the their polar decomposition are all equal to the identity operator. Thus the map $\mathcal{\hat{M}} (x)$ is just $\hat{\Lambda} (x)$. A pure state will evolve by the analog of \eqref{E:Label31}:
\begin{align}
\ket{\psi_t} = \frac{\hat{\Lambda} (x_t) \ket{\psi_0}}{\bra{\psi_0} \hat{\Lambda}^2 (x_t)\ket{\psi_0}^{\frac{1}{2}}}.
\end{align}
As the measurement operators commute, it is easy to take derivatives, and then use the It\"{o} rule to get the quantum diffusion equation for $\ket{\psi_t}$.  In the following we use the Einstein summation rule over repeated upper and lower indices:
\begin{align*}
d\ket{\psi_t} & = \frac{\partial \ket{\psi_t}}{\partial x^i} dx^i + \frac{1}{2} \frac{\partial^2 \ket{\psi_t}}{\partial x^j \partial x^k} dx^j dx^k, \\
\frac{\partial \ket{\psi_t}}{\partial x^i} & = \frac{1}{2}\left( \frac{\hat{M}^2_i}{x^m \hat{M}^2_m} - \frac{\opav{\hat{M}^2_i}_0}{\opav{x^m \hat{M}^2_m}_0} \right) \ket{\psi_t}, \\
\frac{\partial^2 \ket{\psi_t}}{\partial x^j \partial x^k} & = - \frac{1}{4}\left( \frac{\hat{M}^2_k \hat{M}^2_j}{(x^m \hat{M}^2_m)^2} + \frac{\hat{M}^2_j \opav{\hat{M}^2_k}_0}{x^m \hat{M}^2_m \opav{x^l \hat{M}^2_l}_0} \right. \\
& \left. + \frac{\opav{\hat{M}^2_j}_0 \hat{M}^2_k}{\opav{x^m \hat{M}^2_m}_0 x^l \hat{M}^2_l} - 3 \frac{\opav{\hat{M}^2_j}_0 \opav{\hat{M}^2_k}_0}{\opav{x^m \hat{M}^2_m}_0^2} \right) \ket{\psi_t},\\
dx^i & = g^{ij} b_j dt + a_{\alpha}^i dW^{\alpha} = g^{ij} \frac{\opav{\hat{M}^2_j}_0}{\opav{x^m \hat{M}^2_m}_0} dt + a_{\alpha}^i dW^{\alpha}, \\
dx^j dx^k & = g^{jk} dt,
\end{align*}
with $\opav{\hat{M}^2_j}_0 = \bra{\psi_0} \hat{M}^2_j \ket{\psi_0} = \bra{\psi_0} \hat{M}^{\dagger}_j \hat{M}_j \ket{\psi_0}$. It's easy to see that $\frac{\opav{\hat{M}^2_j}_0}{\opav{x^m \hat{M}^2_m}_0} = \opav{\frac{\hat{M}^2_j}{x^m \hat{M}^2_m}}$ where $\opav{\hat{M}^2_j} = \bra{\psi_t} \hat{M}^2_j \ket{\psi_t}$. Denote
\begin{equation}
\hat{A}_i = \frac{\hat{M}^2_i}{x^m \hat{M}^2_m}.
\end{equation}
Putting this all together, we arrive at the following coupled stochastic differential equations:
\begin{align}
d\ket{\psi_t} & = - \frac{1}{8} g^{jk} (\hat{A}_j - \opav{\hat{A}_j}) (\hat{A}_k - \opav{\hat{A}_k}) \ket{\psi_t} dt \notag \\
& \qquad + \frac{1}{2} (\hat{A}_i - \opav{\hat{A}_i}) \ket{\psi_t} a_{\alpha}^i dW^{\alpha}, \notag \\
dx^i & =  g^{ij} \opav{\hat{A}_j} dt + a_{\alpha}^i dW^{\alpha}. \label{E:Label35}
\end{align}
From here we can easily derive an equation for $\rho_t = \ket{\psi_t}\bra{\psi_t}$:
\begin{align}
d\rho_t & = g^{jk} \left( \mathcal{\hat{Q}}_j \rho_t \mathcal{\hat{Q}}_k - \frac{1}{2} \{ \mathcal{\hat{Q}}_k \mathcal{\hat{Q}}_j, \rho_t \} \right) dt + \{ \mathcal{\hat{Q}}_j, \rho_t \} a_{\alpha}^j dW^{\alpha}, \notag \\
dx^i & =  g^{ij} Tr\left( \hat{A}_j \rho_t \right) dt + a_{\alpha}^i dW^{\alpha},
\end{align}
where
\begin{align}
\mathcal{\hat{Q}}_j (\rho_t,x) = \frac{1}{2} \left( \hat{A}_j (x) - Tr \left( \hat{A}_j (x) \rho_t \right) \right).
\end{align}

\section{Conclusions}

In this paper we have constructed stochastic processes that are continuous decompositions of a discrete, instantaneous measurement. The state of the system evolves in accordance with stochastic differential equations, which in the case of generalized measurement involve feedback based on the measurement history. We have proven that in the long-time limit this process arrives at the same final states with the same probabilities as the strong measurement. This gives us the ability to think in terms of this continuous process when we consider measurements, and a natural way to express statements involving measurements in the language of stochastic calculus. For example, we can easily differentiate with respect to the process using It\"{o}'s calculus. One application of this idea has already been made to entanglement monotones, giving new differential conditions for them, as shown in \cite{OrBr1}.

A related and still unsolved question concerns the {\it converse} problem to the one considered in this paper:  namely, if we are able to perform a certain restricted class of operations or if we have control over certain parameters of a quantum system, to what extent we can utilize this freedom to set up an experiment in which the system's evolution is governed by the model considered above?  Given some class of weak measurements that can be experimentally performed, what class of generalized measurements can we generate?  This question has relevance when one starts exploring an experimental realization of the proposed type of continuous measurement, and is the subject of ongoing research.

\section*{Acknowledgments}

We would like to thank Ognyan Oreshkov, Shesha Shayee Raghunathan and Hari Krovi for many useful discussions. This work was supported in part by NSF Grant No.~EMT-0524822, and by a University Grant from Lockheed Martin Corporation.

\end{document}